\pdfoutput=1
\documentclass[12pt]{iopart}
 
\expandafter\let\csname equation*\endcsname\relax
\expandafter\let\csname endequation*\endcsname\relax
\usepackage{amsmath,amssymb}

\usepackage{graphicx,color}
\usepackage{enumerate}

\newcommand\la{\langle}
\newcommand\ra{\rangle}
\newcommand\be{\begin{equation}} 
\newcommand\ee{\end{equation}} 
\begin{document}

\title{
Exact probability distribution for the two-tag displacement  in
single-file motion}

\author{Sanjib Sabhapandit$^1$ and Abhishek Dhar$^2$}  
\address{$^1$Raman Research Institute,  Bangalore - 560080, India\\
$^2$International centre for theoretical sciences, TIFR,
Bangalore - 560012, India}

\date{\today}

\begin{abstract}
We consider a gas of point particles moving on the one-dimensional line with a
hard-core inter-particle interaction that prevents particle crossings --- this is usually referred to as single-file motion.   
 The individual particle dynamics can be arbitrary and they only interact when they meet. 
Starting from initial conditions such that particles are uniformly distributed,  we observe  the
displacement of a tagged particle at time $t$, with respect to the initial
position of \emph{another} tagged particle, such that their tags differ by
$r$. For $r=0$, this is the usual well studied problem of the tagged particle motion. Using a mapping to a non-interacting particle system we compute
 the  exact probability distribution function for the
two-tagged particle displacement, for general single particle
dynamics. As  by-products, we compute the large deviation
function, various cumulants and, for the case of Hamiltonian dynamics,
the two-particle velocity auto-correlation function.
\end{abstract}
\maketitle

\section{Introduction}

Starting with the pioneering work of Jepsen~\cite{jepsen65} and
Harris~\cite{harris65}, the study of tagged particle motion has been
an area of very active research.  Much of the earlier studies
focused \cite{lebowitz67,lebowitz72,percus74,beijeren83,arratia83,pincus78,rodenbeck98,majumdar91,lizana08,barkai09,kollmann2003,gupta07,barkai10,roy13,roy14,sabhapandit:07,illien13,benichou13}
on the statistics of the typical displacement of the tagged particle,
in particular on the mean square displacement in various single-file
systems.  One of the most interesting result is that, for a
Hamiltonian one-dimensional gas of hard particles which move
ballistically between elastic collisions, a tagged particle moves
{\emph{diffusively}}, thus the mean square displacement of the tagged
particle,  in a time duration $t$ increases as $\la
X^2_t\ra \sim t $. On the other hand, for a gas of Brownian particles
with hard-core interactions, the tagged particle motion is
{\emph{sub-diffusive}}, with $\la X_t^2 \ra \propto t^{1/2}$. The
typical fluctuations of the tagged particle displacement is described
by a Gaussian distribution with the above variance. Recently there has
been interest in studying the probability of \emph{atypical} fluctuations of
the tagged particle
displacement \cite{krapivsky:14,hegde:14,krapivsky:15,sadhuderrida:15}.
 Some of the results for these simple
classical interacting particle models have been useful in obtaining
results for one-dimensional quantum systems \cite{damle05,rapp06}.

A number of different theoretical approaches have been used to study
the probability distribution of the tagged particle
displacement. These include the original ideas of Jepsen and Harris of
mapping to a non-interacting system, exact solution of multi-particle
Fokker-Planck equation with reflecting boundary condition between
neighboring particles, and the recently developed approach of Macroscopic
fluctuation theory.  In our recent works \cite{roy13,hegde:14}, we
have shown a simpler way (as opposed to earlier approaches
in \cite{jepsen65,lebowitz67,lebowitz72}) of using the non-interacting
picture to computing tagged particle statistics. In the present work,
we extend this method to study a particular two particle distribution,
defined below.

We consider a hard-point particle system on the infinite line. The
particles are distributed uniformly with a finite density $\rho$. For
the case of Hamiltonian dynamics (of equal mass particles), the
initial velocities are taken to be independent and identically
distributed random variables. The particles move ballistically in
between elastic binary collisions. During collisions, the two
colliding particles merely interchange their velocities. As a result,
any set of particle trajectories of the interacting system can be
constructed from the set of trajectories of a non-interacting system
 (where the trajectories pass through each other) --- by
exchanging the identities (tags) of the  particles at
crossing. For the case of Brownian particles with hard-point
interactions, Harris \cite{harris65} defined the interacting particle
problem by starting with the non-interacting trajectories and
exchanging particle identities whenever two trajectories cross. This
definition is equivalent to enforcing reflecting boundary conditions
between nearest neighbor pairs in the full multi-particle propagator
(thus each particle acts like a hard reflecting wall for its nearest
neighbors).

One can generalize this collision rule to other cases where the
individual particle dynamics is neither Hamiltonian or Brownian, for
example L\`evy walks or fractional Brownian motion.  Hence in general
we define the interacting problem as follows: start with the
non-interacting trajectories and interchange particle labels whenever
two trajectories cross.

For this general interacting particle system, let us tag two particles
with tag indices $0$ and $r$, such that there are $r-1$ particles in
between them.  Let the position of these particles be $x_0(t)$ and
$x_r(t)$ respectively. Here we consider the displacement
$X_t=x_r(t)-x_0(0)$ and compute it's statistics.

\section{Main steps of the calculation}

\begin{figure}
\centering
\includegraphics[width=.5\hsize]{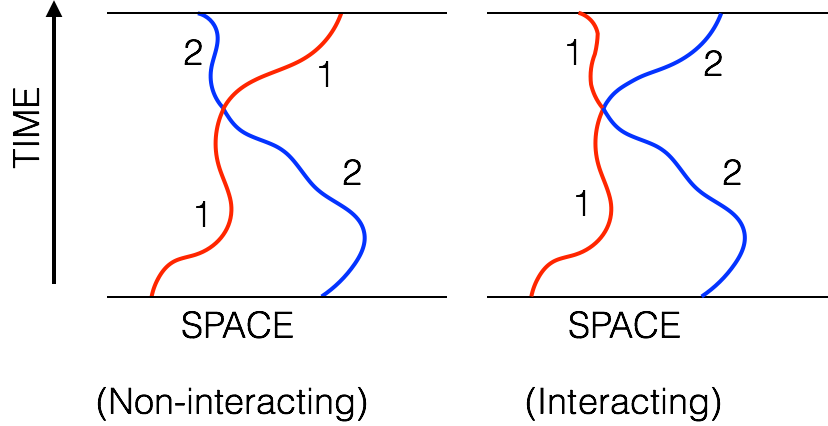}
\caption{\label{tag-exchange} (Color online) An interacting hard-point
particles system can be constructed from a non-interacting system by
exchanging tags (colors) when two trajectories cross.}
\end{figure}

Initially, we consider $2N + 1$ particles, independently and uniformly
distributed in the interval $[-L, L]$ and evolve them on the infinite
one dimensional line.  Since during a collision each particle acts as
a reflecting hard wall for the other and the particles are identical,
one can effectively treat the system of the interacting hard-point
particles as non-interacting by exchanging the identities of the
particles emerging from collisions [see \fref{tag-exchange} and
discussion in previous section].  In the non-interacting picture, each
particle executes an independent motion and the particles \emph{pass
through each other} when they `collide'.  The position of each
particle at time $t$ is given independently by a single-particle
propagator of the general form
\begin{equation}
G(y,t|x,0)=
\frac{1}{\sigma_t}\,f\left(\frac{y-x}{\sigma_t}\right),
\label{propagator} 
\end{equation}
where $f(-w)=f(w) \ge 0$ and $\langle
|y-x|\rangle/\sigma_t=\int_{-\infty}^\infty |w| f(w)\, dw=\Delta$ is
finite. Evidently, $\int_{-\infty}^\infty f(w)\, dw=1$. 
The dependence on time only appears through the characteristic
displacement $\sigma_t$ in time $t$.  While for stochastic processes
the propagator arises naturally, for Hamiltonian systems (where the
dynamics is deterministic) it comes from the 
distribution, taken to be of the form $\bar{v}^{-1} f(v/\bar{v})$,
from which the initial velocities of the particles are chosen
independently.  In many problems of interest, the propagator
happens to be Gaussian, i.e., $f(x)=e^{-x^2/2}/\sqrt{2\pi}$, and
$\sigma_t^2$ is the variance.  For example, for Brownian particles,
$\sigma_t=\sqrt{2 D t}$, where $D$ is the diffusion coefficient, while
for Hamiltonian dynamics with Gaussian velocity distribution we have
$\sigma_t=\bar{v} t$.  Similarly for fractional Brownian motion,
$\sigma_t\propto t^H$, where $H$ is the Hurst exponent. However, our
analysis is valid for any general propagator. Note that the dependence
on time only appears through the characteristic displacement
$\sigma_t$ in time $t$.

\begin{figure}
\centering
\includegraphics[width=.8\hsize]{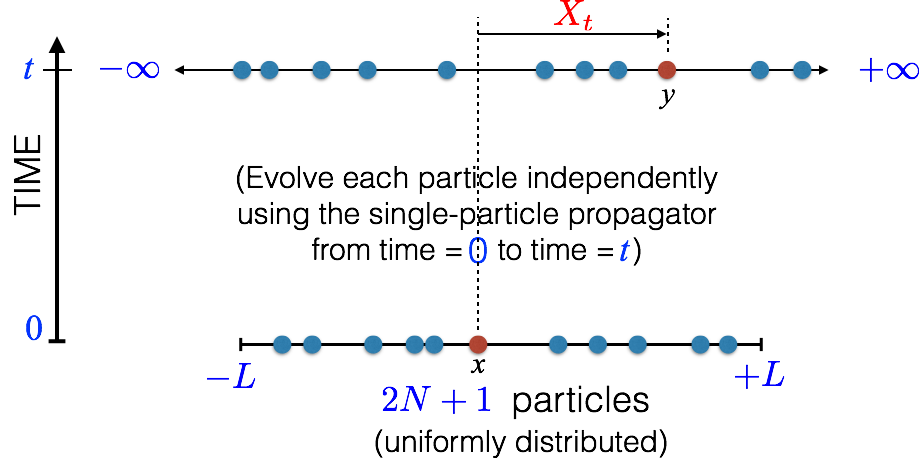}
\caption{\label{evolve} We mark the position  by
$x=x_j(0)$ of the $j^{\rm th}$ particle at $t=0$ and the position
 $y=x_k(t)$ of the $k^{\rm th}$ particle at time
$t$.  In this example, $j=6$ and $k=9$. 
The variable $X_t=y-x$ denotes the displacement of the $k^{\rm th}$
particle at time $t$ with respect to the initial position of the
$j^{\rm th}$ particle.}
\end{figure}

We mark the position $x=x_j(0)$ of the $j^{\rm th}$ particle at $t=0$ and the
position $y=x_k(t)$ of the $k^{\rm th}$ particle at time $t$
[see \fref{evolve}]. Let $X_t=y-x$, be the difference between these two
positions. Our goal is to study the statistical properties of this
random variable $X_t$, in the thermodynamic limit $N \to \infty$,
$L \to \infty$ while keeping $N/L = \rho$ fixed. In this limit, in the
bulk, the statistics of $X_t$ should depend only on the difference of
the tags $r=k-j$, rather than the individual tags $j$ and
$k$. Therefore, we set $j=N+1$ (the middle particle) and $k=N+1+r$
($r^{\rm th}$ particle counted from the middle particle) before taking
the thermodynamic limit. In the thermodynamic limit, $X_t$ denotes the
displacement of a particle at time $t$, with respect to the initial
position of \emph{another} particle such that their tags differ by
$r$. For $r=0$, this represents the usual problem of tagged particle displacement, studied  in \cite{hegde:14}. 

In the following we proceed with the calculation using the
non-interacting picture discussed above.

\subsection{The joint PDF of two particles}

\begin{figure}
\centering
\includegraphics[width=.8\hsize]{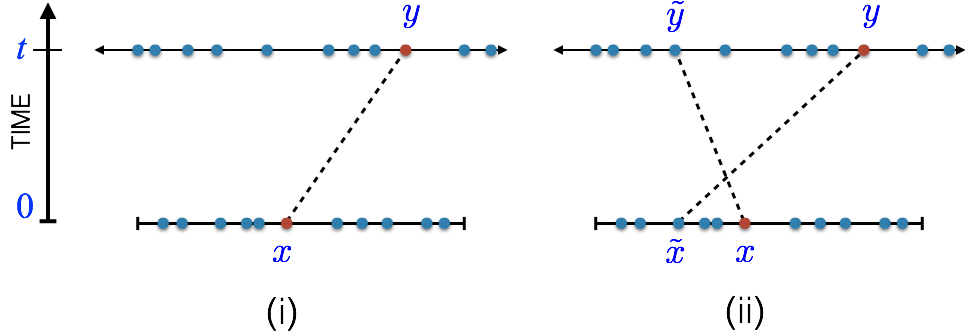}
\caption{\label{two-poss} In the non-interacting
picture, there are two possibilities: (i) the $j^{\rm th}$ particle at
time $t=0$ becomes the $k^{\rm th}$ particle at time $t$, (ii) a
second particle becomes the $k^{\rm th}$ particle at time $t$. For the
case (ii) there are four situations 
(a) $\tilde x <x$ and $\tilde{y} < y$,
(b) $\tilde x >x$ and $\tilde{y} > y$,
(c) $\tilde x <x$ and $\tilde{y} > y$, and
(d) $\tilde x >x$ and $\tilde{y} < y$ respectively.
}
\end{figure}

The joint probability density function (PDF) $P(x,j,0;y,k,t)$ of the
$j^{\rm th}$ particle being at $x$ at time $t=0$, and the $k^{\rm th}$
particle being at $y$ at time $t$, can be expressed in terms of
properties of the non-interacting particles. In the non-interacting
picture, there are two possibilities: (i) the $j^{\rm th}$ particle at
time $t=0$ becomes the $k^{\rm th}$ particle at time $t$, (ii) a
second particle becomes the $k^{\rm th}$ particle at time $t$
[see \fref{two-poss}].  We need to sum over these two processes to
get,
\begin{equation}
P(x,j,0;y,k,t)= P_{(1)}(x,j,0;y,k,t)+ P_{(2)}(x,j,0;y,k,t)~,
\label{ptot}
\end{equation}
where $P_{(1)}$ and $P_{(2)}$ are the joint PDFs corresponding to the
processes (i) and (ii) respectively.

To compute the contribution from process (i) we pick one of the
non-interacting particles at random at time $t=0$, multiply by the
propagator [given in \eref{propagator}] that it goes from $(x, 0)$ to
$(y,t)$, and then multiply by the probability that it is the
$j^{\rm th}$ particle at $t=0$ and the
$k^{\rm th}$ particle at time $t$.  Thus we obtain the
corresponding joint PDF as
\begin{equation}
P_{(1)}(x,j,0;y,k,t)= 
\frac{(2N+1)}{2L} \,G(y,t|x,0)\,F_{1N}(x,j, y, k, t),
\label{P1eq}
\end{equation} 
where $F_{1N}(x, j, y, k, t)$ is the probability that there are
$(j-1)$ particles to the left of $x$ at $t=0$ and $(k-1)$ particles to
the left of $y$ at $t$.

To compute the contribution from process (ii), we first pick two
particles at random at time $t=0$, and multiply by the propagators
that they go from $(x,0)$ to $(\tilde y,t)$ and $(\tilde x, 0)$ to
$(y, t)$ respectively. We then multiply by the probability there are
an $(j-1)$ particles on the left of $x$ at time $t=0$ and $(k-1)$
particles to the left of $y$ at $t$.  Finally, integrating with
respect to $\tilde{x},\tilde{y}$, we get the joint PDF corresponding
to this process as
\begin{equation}
P_{(2)}(x,j,0;y,k,t)=
\frac{(2N+1)(2N)}{(2L)^2}
\int_{-L}^L d\tilde x \int_{-\infty}^\infty 
 d\tilde y \,G(\tilde y,t|x,0)\,G(y,t|\tilde{x},0) \,F_{2N}(x,j,
 y,k, \tilde x,\tilde y, t),    
\label{P2eq}
\end{equation}
where $F_{2N}(x,j,y,k, \tilde x,\tilde y, t)$ is the probability that
there are $(j-1)$ particles on left of $x$ at $t=0$ and $(k-1)$
particles on the left of $ y$ at time $t$, given that there is a
particle at $\tilde x$ at time $t=0$, and a particle at $\tilde y$ at
time $t$.

To proceed further, we need the expressions for $F_{1N}$ and $F_{2N}$.
Let $p_{-+}(x, y,t)$ be the probability that a particle is to the left
of $x$ at $t=0$ and to the right of $y$ at time $t$.  Similarly, we
define the other three complementary probabilities. Clearly,
\begin{subequations}
\begin{align} 
p_{-+}(x,y,t)&=(2L)^{-1}\int_{-L}^x dx' \int_y^{\infty} dy'
G(y',t|x',0),\\
p_{+-}(x,y,t)&=(2L)^{-1}\int_x^{L} dx' \int_{-\infty}^y dy' G(y',t|x',0),\\
p_{--}(x,y,t)&=(2L)^{-1}\int_{-L}^x dx' \int_{-\infty}^y dy' G(y',t|x',0),\\
p_{++}(x,y,t)&=(2L)^{-1}\int_x^{L} dx' \int_y^{\infty} dy' G(y',t|x',0),
\label{transp} 
\end{align} 
\end{subequations}
and $p_{++}+p_{+-} + p_{-+} + p_{--} = 1$. Armed with the above four
probabilities, we now evaluate $F_{1N}$ and $F_{2N}$ below.

\subsection{Evaluation of $F_{1N}(x,j,y,k,t)$} 

\begin{figure}
\centering
\includegraphics[width=.5\hsize]{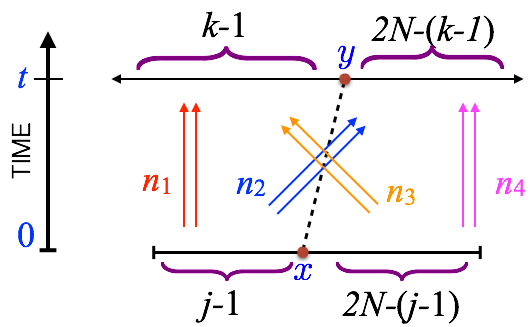}
\caption{\label{figF1N} 
A particle goes from the initial position $x$ to the position $y$ in
time $t$.  $n_1$ denotes the number of particles going from the left
of $x$ to the left of $y$, $n_2$ denotes the number of particles going
from left of $x$ to the right of $y$, $n_3$ denotes the number of
particles going from the right of $x$ to the left of $y$, and $n_4$
denotes the number of particles going from the right of $x$ to the
right of $y$, and $n_1+n_2+n_3+n_4=2N$. There are $j-1$ particles on
the left of $x$ at $t=0$ and $k-1$ particles on the left of $y$ at
time $t$.}
\end{figure}

In this case, out of $2N+1$ particles, the $j^{\rm th}$ particle at
the initial time goes from the position $x$ to the position $y$ in time
$t$ and becomes the $k^{\rm th}$ particle at the final time. The
remaining $2N$ particles are independent of each other and the
selected particle.  Let $n_1$ be the number of particles going from
the left of $x$ to the left of $y$, $n_2$ be the number of particles
going from left of $x$ to the right of $y$, $n_3$ be the number of
particles going from the right of $x$ to the left of $y$, and $n_4$ be
the number of particles going from the right of $x$ to the right of
$y$ [see \fref{figF1N}]. Clearly $n_1+n_2+n_3+n_4=2N$. Moreover, since
there are $j-1$ particles on the left of $x$, clearly, $n_1+n_2 = j-1$
and $n_3 + n_4=2N-(j-1)$. Similarly, since there are $k-1$ particles
on the left of $y$, we have $n_1+n_3=k-1$ and
$n_2+n_4=2N-(k-1)$. These equalities imply $n_4-n_1=2N+2-k-j$ and
$n_3-n_2=k-j$.

The number of ways of choosing the set $\{n_1,n_2,n_3,n_4\}$ is given
by the multinomial coefficient $$\frac{(2N)!}{n_1! n_2! n_3!
n_4!}, $$ and each possibility occurs with probability $ p_{--}^{n_1}
p_{-+}^{n_2} p_{+-}^{n_3} p_{++}^{n_4}.$ Hence, summing over all
possible values of $\{n_1,n_2,n_3,n_4\}$ we get
\begin{equation}
F_{1N} =
\sum_{n_1+n_2+n_3+n_4=2N} \frac{(2N)!}{n_1! n_2! n_3! n_4!} \,
p_{--}^{n_1} p_{-+}^{n_2} p_{+-}^{n_3} p_{++}^{n_4}\,
\delta [n_4-n_1-2N-2+k+j] \,\delta [ n_3-n_2 -k+j],
\label{F1N.1}
\end{equation}
where $\delta[n]$ is the Kronecker delta function: $\delta[n]=1$ if
$n=0$ and $\delta[n]=0$ for $n\not=0$.  Now, after using the integral
representation of the Kronecker delta,
\begin{equation}
\delta [m-n]=\frac{1}{2\pi} \int_{\theta_0}^{\theta_0+2\pi}
e^{i(m-n)\theta} \, d\theta  \quad\text{(where the initial phase
$\theta_0$ is arbitrary)},
\end{equation}
in the above equation, it immediately follows that
\begin{align}
F_{1N}(x,i,y,j,t) =
\int_{\phi_0}^{\phi_0+2\pi} \frac{d\phi}{2\pi}\int_{\theta_0}^{\theta_0+2\pi}\frac{d\theta}{2\pi} 
~\bigl[H(x,y,\theta,\phi,t)\bigr]^{2N} e^{-i \phi (2N+2-k-j)} e^{-i \theta (k-j)},
\end{align}
where
\begin{subequations}
\begin{align}\mspace{-14mu}
\label{H.1}
H(x,y,\theta,\phi,t)&=p_{++}(x,y,t) e^{i\phi} + p_{--}(x,y,t) e^{-i\phi} 
+ p_{+-}(x,y,t) e^{i\theta} + p_{-+}(x,y,t) e^{-i\theta}\\
\label{H.2}
&= 1-(1-\cos{\phi}) ~(p_{++}+ p_{--})+i \sin \phi
~(p_{++}-p_{--}) \notag \\ &~~~\quad -(1-\cos {\theta})~ (p_{+-} +
p_{-+} ) ~ +i \sin \theta~(p_{+-}-p_{-+}) .
\end{align}
\end{subequations}
Now, using the fact that $2N$ is even and the above integral remains
unchanged if both $\phi$ and $\theta$ are shifted by $\pi$
simultaneously, the range of the $\phi$ integral has been broken into
two parts and each of these contributes equally. After appropriately
choosing the initial phases, this gives,
\begin{equation}
F_{1N}(x,y, t) = 
\int_{-\pi/2}^{\pi/2} \frac{d\phi}{\pi}\int_{-\pi}^{\pi}\frac{d\theta}{2\pi} 
\,\bigl[H(x,y,\theta,\phi,t)\bigr]^{2N} 
e^{-i \phi (2N+2-k-j)} e^{-i \theta (k-j)}.
\label{F1eq}
\end{equation}

\subsection{Evaluation of $F_{2N}(x,j,y,k,\tilde{x}, \tilde{y},t)$}

\begin{figure}
\centering
\includegraphics[width=.8\hsize]{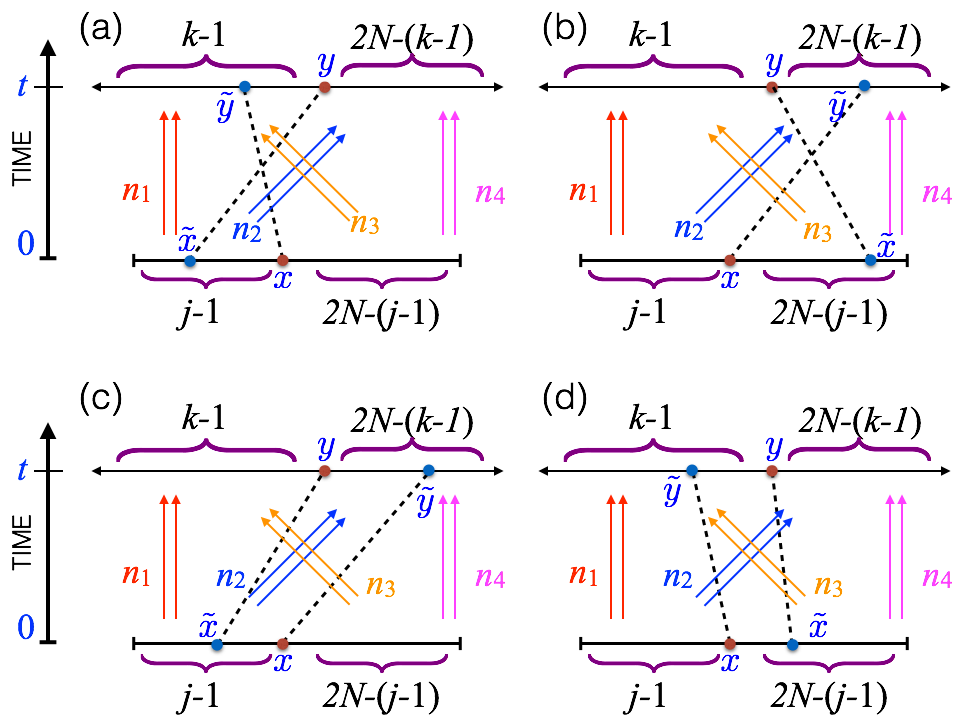}
\caption{\label{figF2N} 
A particle goes from the initial position $x$ to the position
$\tilde{y}$ in time $t$, while at the same time, another particle from
its initial position $\tilde{x}$ goes to $y$.  $n_1$ denotes the
number of particles going from the left of $x$ to the left of $y$,
$n_2$ denotes the number of particles going from left of $x$ to the
right of $y$, $n_3$ denotes the number of particles going from the
right of $x$ to the left of $y$, and $n_4$ denotes the number of
particles going from the right of $x$ to the right of $y$, and
$n_1+n_2+n_3+n_4=2N-1$. There are $j-1$ particles on the left of $x$
at $t=0$ and $k-1$ particles on the left of $y$ at time $t$.}  
\end{figure}

In this case a particle whose initial tag is different from $j$, goes
from $\tilde{x}$ to $y$ and becomes the $k^{\rm th}$ particle at time
$t$, while the initial $j^{\rm th}$ particle goes from $x$ to
$\tilde{y}$ in time $t$ whose final tag is different from $k$. To
compute $F_{2N}$, one has to keep track of both these particles.
Apart from these two particles, let there be $n_1$ particles going
from the left of $x$ to the left of $y$, $n_2$ particles going from
the left of $x$ to the right of $y$, $n_3$ particles going from the
right of $x$ to the left of $y$, and $n_4$ particles going from the
right of $x$ to the right of $y$.  Since two of the particles are
considered separately, the rest can be chosen in $(2N-1)!/(n_1! n_2!
n_3!  n_4!)$ different ways and $n_1+n_2+n_3+n_4=2N-1$. The other two
constraints among $\{n_i\}$'s are given by $n_4-n_1=2N+2-k-j +\chi_1$
and $n_3-n_2=k-j +\chi_2$. Unlike the previous case where
$\chi_1=\chi_2=0$, here their values depend on the order of the
positions $(x,\tilde{x})$ and $(y,\tilde{y})$.  There arises four
situations:
\begin{enumerate}[(a)]

\item  
$\tilde x <x$ and $\tilde{y} < y$, for which 
$\chi_1=1$ and $\chi_2=0$,

\item
$\tilde x >x$ and $\tilde{y} > y$, for which 
$\chi_1=-1$ and $\chi_2=0$,

\item
$\tilde x <x$ and $\tilde{y} > y$, for which
$\chi_1=0$ and $\chi_2=1$, 

\item
$\tilde x >x$ and $\tilde{y} < y$, for which $\chi_1=0$ and
$\chi_2=-1$.

\end{enumerate}

Now following the procedure used to evaluate $F_{1N}$,
it is easily found that
case, one has to keep track of the 
\begin{align}
F_{2N}(x,y,\tilde{x},\tilde{y},t) = 
\int_{-\pi/2}^{\pi/2} \frac{d\phi}{\pi}\int_{-\pi}^{\pi} \frac{d\theta}{2\pi} 
\,&\bigl[H(x,y,\theta,\phi,t)\bigr]^{2N-1} 
e^{-i \phi (2N+2-k-j)} e^{-i \theta (k-j)} e^{-i\phi\chi_1}
e^{-i\theta\chi_2}.
\label{F2eq}
\end{align}

\subsection{Exact PDF of the two-tag displacement in the thermodynamic limit}

So far our calculations are exact, valid for any $N$ and $L$.  We now
assume both $N$ and $L$ to be large and keep only the dominant
terms. Finally, we will take the thermodynamic limit $N\to\infty$,
$L\to\infty$ while keeping $N/L=\rho$ fixed.  We now set $k=j+r$ and
$j=N+1$. We also change our notation to
$P_{(1)}(x,N+1,0;y,N+1+r,t) \to P_{(1)}(x,y,r,t)$,
$P_{(2)}(x,N+1,0;y,N+1+r,t) \to P_{(2)}(x,y,r,t)$ and
\begin{equation}
P(x,y,r,t)= P_{(1)}(x,y,r,t) + P_{(2)}(x,y,r,t).
\label{Pfinal}
\end{equation}

From \eref{P1eq} and \eref{F1eq} we get
\begin{equation}
P_{(1)}(x,y,r,t)=\frac{\rho}{\sigma_t} f(z)\, 
\int_{-\pi/2}^{\pi/2} \frac{d\phi}{\pi}\int_{-\pi}^{\pi}\frac{d\theta}{2\pi} 
\,\bigl[H(x,y,\theta,\phi,t)\bigr]^{2N} 
e^{i \phi r} e^{-i \theta r},
\label{P1final}
\end{equation}
where $z=(y-x)/\sigma_t$.
Similarly from \eref{P2eq} and \eref{F2eq} and performing the integration over $\tilde{x}$ and $\tilde{y}$, we get 
\begin{align}
P_{(2)}(x,y,r,t)=& 
\rho^2 \int_{-\pi/2}^{\pi/2} \frac{d\phi}{\pi}\int_{-\pi}^{\pi}\frac{d\theta}{2\pi} 
\,\bigl[H(x,y,\theta,\phi,t)\bigr]^{2N-1} e^{i \phi r} e^{-i \theta r}\notag\\
&\qquad\times\bigl[2 A_1(z) A_2(z) \cos\phi + A_1^2(z) e^{-i\theta} +
  A_2^2(z)e^{i\theta}\bigr] ,
\label{P2final}
\end{align}
where the functions $A_{1,2}(z)$ are given by
\begin{align}
A_1(z)=\int_{\sigma_t z}^\infty G(x,t|0,0)\, dx=\int_z^\infty f(w)\, dw,
~~~\text{and}~~A_2(z)&=1-A_1(z).
\end{align}

Now we explicitly compute the expressions for $p_{\pm\pm}$ using
\eref{propagator}. Keeping only the dominant terms up to  $O(1/L)$,
which survive in the limit $N\to \infty,~L \to \infty$ while keeping
$N/L=\rho$ fixed, we get
\begin{subequations}
\begin{align}
p_{-+}&=\frac{\sigma_t}{2L}\left[-\frac{z}{2} + Q(z) \right] +\dotsb\\
  p_{+-}&=\frac{\sigma_t}{2L}\left[\frac{z}{2} + Q(z) \right]
  +\dotsb\\ p_{--}&=\frac{1}{2}
  +\frac{\sigma_t}{2L}\left[\frac{\bar{z}}{2} - Q(z) \right]
  +\dotsb \\ p_{++}&=\frac{1}{2}
  +\frac{\sigma_t}{2L}\left[-\frac{\bar{z}}{2} - Q(z) \right] +\dotsb,
\end{align}
\end{subequations}
where  $z=(y- x)/\sigma_t$, $\bar{z}=(y+ x)/\sigma_t$, and 
\begin{equation}
\label{Q(z)}
Q(z)= z\int_0^z f(w)\, dw + \int_z^\infty w f(w)\, dw.
\end{equation}

Now, substituting $p_{\pm\pm}$ in the  expression \eref{H.2}, for
large $N$, keeping only the most dominant terms, one finds
\begin{equation}
H^{2N}= e^{-2N (1-\cos\phi)}
e^{-i  \rho \sigma_t \bar{z} \sin \phi}
e^{-2 \rho\sigma_t Q(z) (1-\cos \theta)} e^{i \rho\sigma_t z \sin \theta}.
\label{H.final}
\end{equation}

The PDF of $X_t=y-x$ is given by
\begin{equation}
P_\text{tag}(X_t,r,t)=\int_{-\infty}^\infty\int_{-\infty}^\infty \delta\bigl(X_t -[y-x] \bigr)\,
P(x,y,r,t)\, dx\, dy.
\end{equation}
Using \eref{H.final}, in \eref{Pfinal}--\eref{P2final}, and making a
change of variables from $x,y$ to $z, \bar{z}$, we can finally write
down the above PDF of $X_t$ as
\begin{align}
P_\text{tag}(X_t=\sigma_tz,r,t)=&
\lim_{N\to\infty} \int_{-\infty}^\infty \frac{d\bar{z}}{2}
\int_{-\pi/2}^{\pi/2}
\frac{d\phi}{\pi}\int_{-\pi}^{\pi}\frac{d\theta}{2\pi} 
\rho B(z,\theta,\phi)\notag\\
&\times e^{-2N (1-\cos\phi)}
e^{-i  \rho \sigma_t \bar{z} \sin \phi}
e^{-2 \rho\sigma_t Q(z) (1-\cos \theta)} e^{i \rho\sigma_t z \sin \theta} \,
e^{i \phi r} e^{-i \theta r},
\end{align} 
where 
\begin{math}
B(z,\theta,\phi)=f(z) +\rho\sigma_t
\bigl[2 A_1(z) A_2(z) \cos\phi + A_1^2(z) e^{-i\theta} +
  A_2^2(z)e^{i\theta}\bigr] .
\end{math}
For large $N$, the major contribution of the integral over $\phi$
comes from the region around $\phi=0$. Therefore, the $\phi$ integral
can be performed by expanding around $\phi=0$ to make it a Gaussian
integral (while extending the limits to $\pm\infty$). Subsequently,
one can also perform the Gaussian integral over $\bar{z}$. This leads
to the exact expression, 
\begin{equation}
P_\text{tag}(X_t=\sigma_t z,r,t)=
\frac{1}{\sigma_t}
\int_{-\pi}^{\pi}\frac{d\theta}{2\pi} \,
B(z,\theta)
\,e^{-\rho\sigma_t \bigl[2  Q(z) (1-\cos \theta)-i  z \sin \theta\bigr]}\,
e^{-i \theta r},
\label{P_exact}
\end{equation}
where
\begin{equation}
\label{B(z)}
B(z,\theta)\equiv B(z,\theta,0)
 = f(z)+
\rho\sigma_t
\bigl[2 A_1(z) A_2(z) + A_1^2(z) e^{-i\theta} +
  A_2^2(z)e^{i\theta}\bigr].
\end{equation}

In fact, using the integral representation of the modified Bessel
function of the first kind, for integer order $n$,
\begin{equation}
I_n(x)= \int_{-\pi}^{\pi} \frac{d\theta}{2\pi} e^{x \cos\theta} \,
e^{ \pm i n \theta}, 
\qquad\bigl[\text{evidently, }~ I_{-n}(x)= I_n(x)\bigr]
\end{equation}
the above exact PDF given by \eref{P_exact} can be expressed in the
closed form,
\begin{align}
P_\text{tag}(X_t=\sigma_t
z,r,t)&= \frac{1}{\sigma_t}\,
e^{-2\rho\sigma_t Q(z)}
\left[
\frac{\sqrt{2Q(z)+z}}{\sqrt{2Q(z)-z}}
\right]^{r}\notag\\
&\times\,
\Biggl\{
\Bigl[f(z) + 2 \rho\sigma_t\, A_1(z) A_2(z)\Bigr]
I_r\Bigl(\rho\sigma_t\,\sqrt{4Q^2(z) -z^2}\Bigr) \notag\\
&\qquad+ \rho\sigma_t\, A_1^2(z) \,\frac{\sqrt{2Q(z)+z}}{\sqrt{2Q(z)-z}}\,I_{r+1}\Bigl(\rho\sigma_t\sqrt{4Q^2(z)-z^2}\Bigr) \notag\\
&\qquad+ \rho\sigma_t \,A_2^2(z) \frac{\sqrt{2Q(z)-z}}{\sqrt{2Q(z)+z}} I_{r-1}\Bigl(\rho\sigma_t\sqrt{4Q^2(z) -z^2}\Bigr) 
\Biggr\}. \label{Pexact}
\end{align}

\section{Large deviation result for the two-tags displacement}

In this section, we obtain the large deviation form of \eref{P_exact}
by evaluating the $\theta$ integral using saddle point
approximation. Note that in the expression of $B(\theta,z)$
in \eref{B(z)}, the second term is larger by $O(\rho\sigma_t)$
compared to the first term $f(z)$, which comes from the process (i)
where, in the non-interacting picture, the same particle happens to be
the $j^{\rm th}$ and $(j+r)^{\rm th}$ particles at the initial and
final times respectively. So this process does not contribute at
$O(\rho\sigma_t)$.

In this case we also scale the final with $\rho\sigma_t$. The saddle
point approximation of the integral in \eref{P_exact} gives
\begin{equation}
P_\text{tag}(X_t=\sigma_t z,r=\rho\sigma_tl,t) \approx
\frac{1}{\sigma_t}
\frac{\sqrt{\rho\sigma_t}}{\sqrt{2\pi g_2(z)}}\,
g_1(z)\,e^{-\rho\sigma_t I(z)} ,
\label{analytic}
\end{equation}
where the different functions are explained below.

The large deviation function (rate function) is given by
\begin{equation}
I(z)= 2 Q(z)(1-\cos \theta^*) -i z \sin \theta^* +i \theta^* l,
\label{I.1}
\end{equation}
where the saddle point $\theta^*$ is obtained using the condition
\begin{equation}
\frac{\partial}{\partial \theta}
\bigl[2  Q(z) (1-\cos \theta)-i  z \sin \theta +i\theta l
\bigr]\Big|_{\theta=\theta^*}=0,
\end{equation}
which gives
\begin{equation}
e^{\pm i\theta^*}=\frac{\pm l+\sqrt{l^2+4Q^2(z)-z^2}}{2Q(z)\pm z}.
\label{theta*}
\end{equation}
Substituting $\theta^*$ in \eref{I.1} gives the large deviation
function, explicitly in terms of $z$ as
\begin{equation}
I(z)=2Q(z)-\sqrt{l^2+4Q^2(z)-z^2} +
l\ln \left[\frac{l+\sqrt{l^2+4Q^2(z)-z^2}}{2Q(z)+z}\right].
\label{I(z)}
\end{equation}

The function $g_1(z)$ comes from evaluating the prefactor in \eref{P_exact}
at the saddle point,
\begin{align}
g_1(z)&=  (\rho\sigma_t)^{-1} \, B(z,\theta^*)\notag\\
&=\Bigl[2 A_1(z) A_2(z) + A_1^2(z) e^{-i\theta^*} +
  A_2^2(z)e^{i\theta^*}\Bigr]
+ O\bigl([\rho\sigma_t]^{-1}\bigr),
\label{gz}
\end{align}
where $e^{\pm i \theta^*}$ are given in \eref{theta*}.

The function $g_2(z)$ comes from performing the Gaussian integral
around the saddle-point $\theta^*$,
\begin{equation}
g_2(z)=\frac{\partial^2}{\partial \theta^2}
\bigl[2  Q(z) (1-\cos \theta)-i  z \sin \theta + i\theta
l\bigr]\Big|_{\theta=\theta^*} 
= \sqrt{l^2+ 4 Q^2(z)-z^2}. 
\end{equation}

\begin{figure}
\centering
\includegraphics[width=.8\hsize]{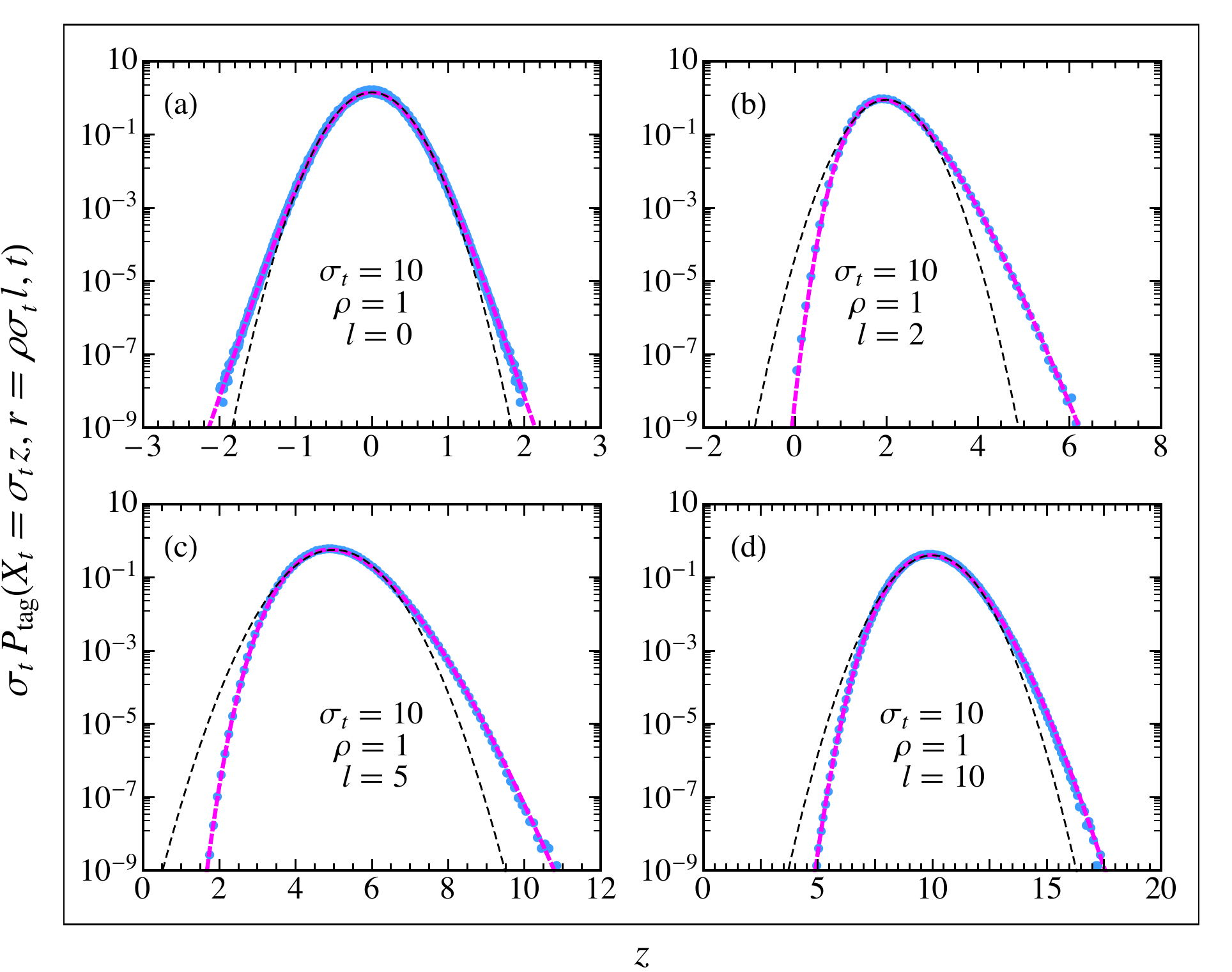} 
\caption{(Color online)
The (blue) points represent the simulation results for the PDF of the
two-tag displacement in a one-dimensional system of hard-point
particles for tag separations (a) $r=0$, (b) $r=20$, (c) $r=50$, and
(d) $r=100$ respectively.  The (magenta) thick dashed line corresponds
to the analytic result in Eq.~(\ref{analytic}), while the (black) thin
dashed line is for Gaussian distribution with the variance given
by \eref{c2}.  }
\label{pdf}
\end{figure}

The large deviation function has a minimum at $z=l$ and near this
minimum, we get 
\begin{equation}
I(z)=\frac{1}{2}\frac{(z-l)^2}{2Q(l)} + O\bigl([z-l]^3\bigr).
\end{equation}
Therefore, near the peak at $X_t=r/\rho$ the PDF
$P_\text{tag}(X_t,r,t)$ has a Gaussian form, which describes the
\emph{typical} fluctuations. However, away from this central region,
the Gaussian approximation breaks down, and one require the large
 deviation result \eref{analytic} to describe for the \emph{atypical}
 large fluctuations. In \fref{pdf} we compare both the Gaussian
 approximation and the large deviation result of the PDF with
 numerical simulation, and find that while the Gaussian approximation
 fits the data well near the central peak, the large deviation result
 agrees very well with the simulation data even beyond the central
 region.

\section{Cumulants}

Now, we look at the cumulant generating function of the two-tag
 displacement $X_t$. We define
\begin{equation}
Z(\lambda)
=\Bigl\langle e^{\lambda \rho X_t}\Bigr\rangle
=e^{\rho \sigma_t \mu(\lambda)}, 
\end{equation}
such that the expansion of $\mu(\lambda)$ in terms of the cumulants is
given by
\begin{equation}
\mu(\lambda)
= \frac{1}{\rho \sigma_t}\sum_{n=1}^\infty \frac{(\lambda\rho)^n}{n!}\, \langle
X_t^n \rangle_c ~.
\label{mu-expansion}
\end{equation}

Using the large deviation form of $P_\mathrm{tag}(X_t,r,t)$ given
by \eref{analytic}, and then evaluating the integral over $z$ using the
saddle point approximation, we have $\mu(\lambda)=\lambda z^* -
I(z^*)$ where $z^*$ is implicitly given by the equation
$\lambda=dI(z^*)/dz^*$. Using the expression of $I(z)$ obtained above in
terms of $\theta^*$ with the substitution 
$e^{-i\theta^*}=\nu$
we can
express $\mu(\lambda)$ in the parametric form
\begin{subequations}
\label{mu-parametric}
\begin{align}
\label{mu}
&\mu(\lambda)= \lambda z + \frac{1-\nu}{1+\nu} (z+l)+  l\ln \nu, \\
\label{lambda}
&\lambda=\bigl(1-\nu^{-1}\bigr)
\left[1+ (\nu-1)\,A_1(z) \right], \\
\label{B}
&\nu=\frac{- l+\sqrt{l^2+4Q^2(z)-z^2}}{2Q(z)- z}. 
\end{align}
\end{subequations}

The absence of $\rho$ and $\sigma$ in \eref{mu-parametric} indicates
that $\mu(\lambda)$ does not depend on them. Therefore,
from \eref{mu-expansion}, it follows that 
\begin{equation}
\langle
X_t^n \rangle_c \propto \frac{\sigma_t}{\rho^{n-1}}.
\end{equation}

Note that the above equations \eref{mu-parametric}, obtained through
the saddle point calculation, gives the cumulants only at the most
dominant order $O(\sigma_t)$.  To obtain the cumulants, we first
expand the right hand side of \eref{B} about $z=l$ and then invert the
series to obtain $z$ in terms of a series in $\nu$ about
$\nu=1$. Therefore, the right hand side of \eref{lambda} can be
expressed as a series in $(\nu-1)$.  Next, by inverting \eref{lambda},
we obtain $\nu$ (and hence also $z$) in terms of a series in $\lambda$
about $\lambda=0$. Finally, from \eref{mu}, we express $\mu(\lambda)$
as a series in $\lambda$, and using the definition
in \eref{mu-expansion}, we obtain the first few cumulants (mean,
variance, skewness, and kurtosis respectively) as
\begin{subequations}
\begin{align}
\label{c1}
\sigma_t^{-1}\langle X_t \rangle_c &=l, \\
\label{c2}
\sigma_t^{-1}\rho\langle X_t^2 \rangle_c &=2 Q(l)\\
\label{c3}
\sigma_t^{-1}\rho^2\langle X_t^3 \rangle_c &=
12 F(l) Q(l)-l, 
\\
\label{c4}
\sigma_t^{-1}\rho^3\langle X_t^4 \rangle_c &=
6 \Bigl[8 f(l) Q^2(l)-Q(l) + 20 F^2(l) Q(l)-2 l F(l)\bigr], 
\end{align}
\end{subequations}
where $l=r/(\rho\sigma_t)$ and 
\begin{math}
F(l) = \int_0^l f(w)\, dw.
\end{math}

While the mean $\langle X_t \rangle_c =r/\rho$ is exact to all order,
the expressions for the other three cumulants are exact only at the
leading order $O(\sigma_t)$. The sub-dominant corrections can be
computed using the exact expression of the PDF given
by \eref{P_exact}. For example, for the variance we get
\begin{align}
\langle X_t^2 \rangle_c &=
\frac{1}{\rho^2} \Biggl\{
2  \rho\sigma_t\, Q(l) +
\biggl[2 f(l) Q(l)+2 F^2(l)-\frac{1}{2}\biggr]\notag\\
&+
[\rho\sigma_t]^{-1}\biggl[
Q^2(l) f''(l)+f'(l) 
\Bigl(8 F(l) Q(l)-\frac{l}{3}\Bigr)+f(l) 
\Bigl(12 F^2(l)-1\Bigr)+6 f^2(l) Q(l)
\biggr]
 \notag \\
&+O\bigl([\rho\sigma_t]^{-2}\bigr)
\Biggr\}. \label{varexp}
\end{align}

\section{Velocity autocorrelations}
As a spin-off of our calculation, we show here that for the
Hamiltonian model of elastically colliding particles, we can also
compute the velocity auto-correlation function $\la v_0(0) v_r
(t) \ra$. This can in fact be derived directly from the positional
correlation function. We note that for the Hamiltonian case, we have
$x_i(t)= \int_0^t dt' v_i(t')$. Hence it follows that
\begin{equation}
\frac{1}{2}\frac{d}{dt} \la [x_r(t)-x_0(0)]^2 \ra = 
\frac{1}{2}\frac{d}{dt} \Bigl[ \la x_r^2(t) \ra + \la x_0^2(0) \ra - 2 \la x_r(t)x_0(0) \ra \Bigr]~.
\end{equation}
The first two terms inside the square bracket are 
independent of time, hence they drop off on taking a
time-derivative. For a Hamiltonian system we have $(d/dt)\la x_r(t)
x_0(0)\ra = \la v_r(t) x_0(0)\ra= \la v_r(0) x_0(-t)\ra$.  Taking
another derivative, we get
\begin{equation}
\frac{1}{2}\frac{d^2}{dt^2} \bigl\la [x_r(t)-x_0(0)]^2 \bigr\ra = \la v_r(t) v_0(0)\ra~.
\end{equation}  
Using Eq.~(\ref{varexp}), the fact that $d^2 Q/dl^2=f(l)$, and
$\sigma_t=\bar{v} t$ for Hamiltonian dynamics, we therefore get
\begin{align}
\frac{1}{\bar{v}^2}\, \la v_r(t) v_0(0)\ra &= \frac{1}{2}\frac{d^2}{d\sigma_t^2}~\langle X_t^2 \rangle_c \label{vaf} \\ &= \frac{1}{\rho \sigma_t} l^2 f(l) + O\left([\rho \sigma_t]^{-2}\right). \label{vafLO}
\end{align}
\begin{figure}
\centering
\includegraphics[width=.8\hsize]{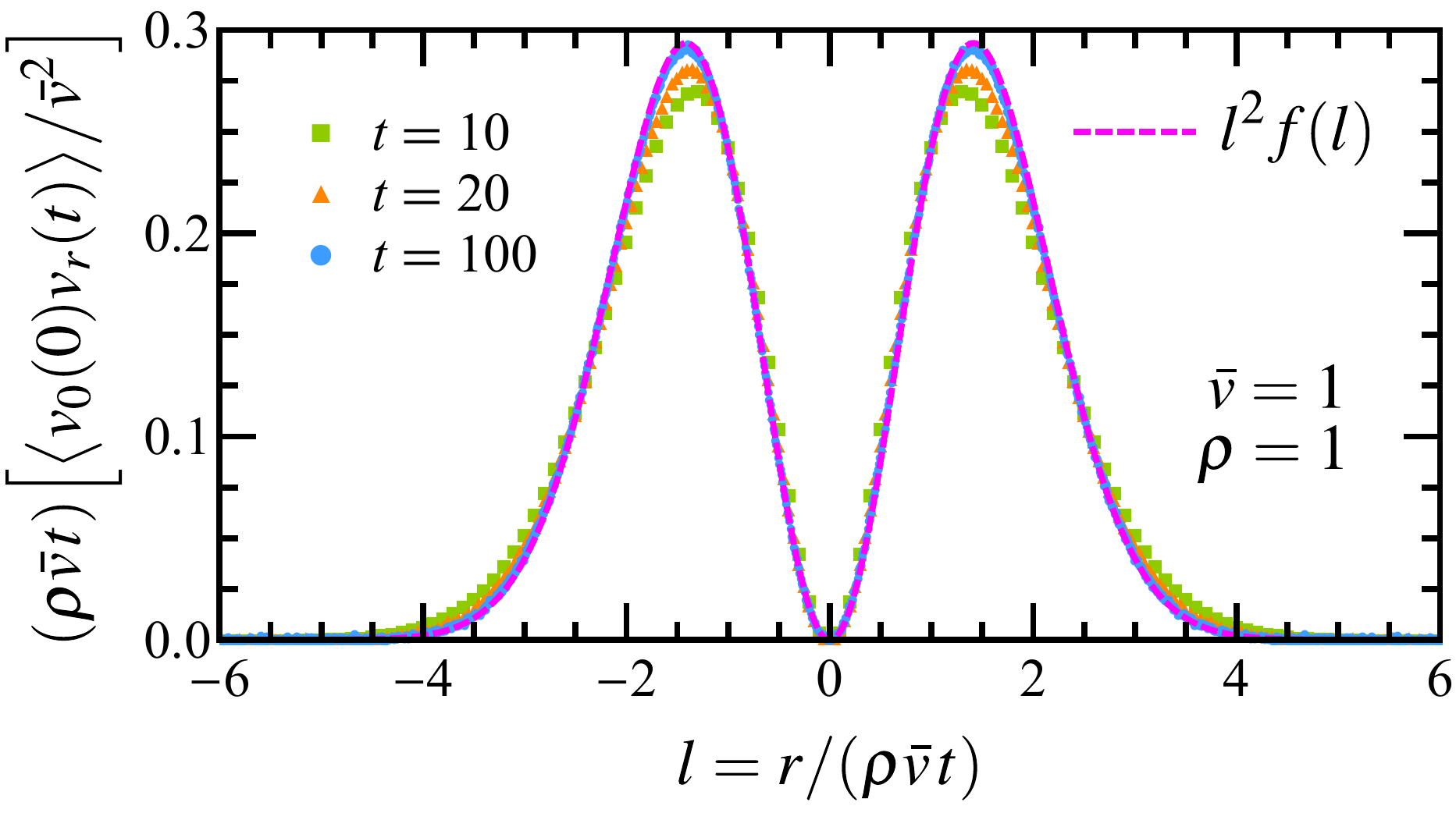}
\caption{\label{velocity-correlation} The points are numerical
simulation results for the velocity auto-correlation function computed
from two tagged particles as a function of the tag separation at three
different times, for the Hamiltonian model. The dashed line represent
the analytical result to the leading order.}
\end{figure}

To leading order, this result can be obtained by a simple
argument.  Since the initial velocities are chosen
independently for each particle, the contribution to the correlation
function $\la v_r(t) v_0(0)\ra$ is non-zero only when the velocity of
the $r-$th particle at time $t$ is the same as that of the zero-th
particle at time $t$. Thus we have
\begin{align}
\bigl\langle v_0(0) v_r(t)\bigr\rangle 
\simeq \bigl\langle \delta (r-\rho v t) \,v^2\bigr\rangle 
&= \frac{1}{\rho t} \left\langle \delta \left(v-\frac{r}{\rho t}\right) 
\,v^2\right\rangle \notag\\
&=  \frac{1}{\rho t} \int \delta \left(v-\frac{r}{\rho t}\right) 
\,v^2 \,\frac{1}{\bar{v}} f\left(\frac{v}{\bar{v}}\right)\, dv~,
\end{align}
hence finally 
\begin{equation}
\frac{1}{\bar{v}^2}\, \bigl\langle v_0(0) v_r(t)\bigr\rangle 
\simeq \frac{1}{\rho \bar{v} t} \left(\frac{r}{\rho\bar{v} t} \right)^2
f\left(\frac{r}{\rho\bar{v} t}\right),
\end{equation}
as in Eq.~(\ref{vafLO}). In \fref{velocity-correlation} we show a
comparison of this analytic result with direct simulation results for
the two-particle velocity autocorrelations in the equilibrium
hard-particle gas.  An exact expression for the velocity
autocorrelation was obtained in \cite{jepsen65} and involves a very
lengthy calculation. This exact result can be recovered from using
Eqs.~(\ref{Pexact},\ref{vaf}). However we see that the leading order
expression is already quite accurate in describing the long time
behavior.   For the special case $r=0$ and equilibrium initial
conditions, using \eref{varexp} in \eref{vaf}, we get
\begin{equation}
\frac{1}{\bar{v}^2}\, \bigl\langle v_0(0) v_r(t)\bigr\rangle \simeq
(\rho\bar{v} t)^{-3} \Bigl[Q^2(0) f''(0)
-f(0) + 6 Q(0) f^2(0) 
\Bigr].
\end{equation}
For the Gaussian distribution, $f(x)=\exp(-x^2/2)/\sqrt{2\pi}$, this
gives $\la v_0(t) v_0(0)\ra/\bar{v}^2 \simeq -(\rho\bar{v}t)^{-3}
(2 \pi-5)/(2\pi)^{3/2} $, a result first derived in \cite{jepsen65},
and also in \cite{roy13} using the present approach.  
\section{Discussion}
 We have considered a system of point particles moving on
a one-dimensional line. The dynamics of individual particles is
arbitrary and can be either stochastic or deterministic, and the only
interaction between the particles is when they meet. The interaction
dynamics is specified by imposing that, we start with the
non-interacting trajectories, and then interchange particle labels
whenever trajectories cross --- thus the ordering of particle labels
is maintained at all times. This dynamics is quite natural for the
deterministic so-called Jepsen gas and also for non-crossing Brownian
walkers, and also seems natural for other stochastic processes.  Using
the fact that a mapping to non-interacting particles is available we
have developed a formalism that seems to be very suited to computing
reduced distribution functions and correlation functions in the
interacting system. In particular, here we focus on computing the
joint distribution of the positions of two tagged particles at
different times. This is obtained exactly and from this we extract the
large deviation function, various cumulants. For the case of
Hamiltonian dynamics, we show that the two-pont velocity
autocorrelation function can also be computed. We expect that the
general strategy of our approach will be useful in the computation of
more complicated correlations and distribution functions.

\section{Acknowledgments} We thank the Galileo Galilei Institute for Theoretical Physics for the hospitality and the INFN for partial support during the completion of this work.

\end{document}